\documentclass[aps,prc,twocolumn,superscriptaddress,nofootinbib,showpacs]{revtex4-1}
\usepackage{graphicx,amsmath,amssymb,bm}

\newcommand{\bnu}{\bar{\nu}}
\newcommand{\calM}{{\cal M}}
\newcommand{\calT}{{\cal T}}
\newcommand{\mred}{m_{\rm red}}
\newcommand{\sw}{\rm S}
\newcommand{\pw}{\rm P}
\newcommand{\dw}{\rm D}

\begin{document}

\title{Neutrino-pair bremsstrahlung from nucleon-$\alpha$ versus nucleon-nucleon scattering}

\author{Rishi Sharma}
\affiliation{TIFR, Homi Bhabha Road, Navy Nagar, Mumbai 400005, India}
\affiliation{TRIUMF, 4004 Wesbrook Mall, Vancouver, BC, V6T 2A3, Canada}
\author{Sonia Bacca}
\affiliation{TRIUMF, 4004 Wesbrook Mall, Vancouver, BC, V6T 2A3, Canada}
\affiliation{Department of Physics and Astronomy, University of Manitoba, Winnipeg, MB, R3T 2N2, Canada}
\author{A.\ Schwenk}
\affiliation{Institut f\"{u}r Kernphysik,
Technische Universit\"{a}t Darmstadt, 64289 Darmstadt, Germany}
\affiliation{ExtreMe Matter Institute EMMI,
GSI Helmholtzzentrum f\"{u}r Schwerionenforschung GmbH,
64291 Darmstadt, Germany}

\begin{abstract}
We study the impact of the nucleon-$\alpha$ P-wave resonances on
neutrino-pair bremsstrahlung. Because of the non-central spin-orbit
interaction, these resonances lead to an enhanced contribution to the
nucleon spin structure factor for temperatures $T \lesssim 4$~MeV.
If the $\alpha$-particle fraction is significant and the temperature is
in this range, this contribution is competitive with neutron-neutron
bremsstrahlung. This may be relevant for neutrino production in
core-collapse supernovae or other dense astrophysical
environments. Similar enhancements are expected for resonant
non-central nucleon-nucleus interactions.
\end{abstract}

\pacs{97.60.Bw, 26.50.+x, 26.60.-c, 95.30.Cq}

\preprint{TIFR/TH/14-22}

\maketitle

\section{Introduction}

Neutrinos provide a window to the ``core'' of supernova explosions.
Due to their weak interactions, they are liberated seconds after the
core bounce and take away $99\%$ of the gravitational energy from the
core collapse, undergoing interactions with the surrounding
matter~\cite{RaffeltBook,Burrows:2004} and with each
other~\cite{Qian:1994,Pastor:2002}. (For recent reviews on core-collapse
supernova explosions, see Refs.~\cite{Janka:2012,Burrows:2013}.)
Neutrinos play a role in the revival of the shock, and their arrival
timing and spectrum can tell us about the physical conditions and the
explosion dynamics. Moreover, neutrinos from the proto-neutron star
provide an important source for
nucleosynthesis~\cite{Janka:2006,Duan:2011,Arcones:2012}.

Therefore, it is important to identify and quantitatively determine
the neutrino production, scattering, and absorption mechanisms for the
relevant astrophysical conditions. Many different leptonic and
hadronic processes of neutrino production have been considered in the
literature (see, e.g., Refs.~\cite{RaffeltBook,Burrows:2004,Janka:2012}).
In this paper, we will focus on the emission of neutrino-antineutrino
($\nu\bnu$) pairs by hadronic bremsstrahlung processes. These provide
an important source of neutrino production, in particular for $\mu$
and $\tau$ neutrinos~\cite{Hannestad:1998,Thompson:2000} which are
not generated by charged-current reactions.

For neutrino processes involving strongly interacting matter, it is
convenient to write these in terms of the structure factor or the
response function. For bremsstrahlung, the relevant one is the spin
structure factor. This is because non-central nuclear interactions do
not conserve spin, so there is a non-zero spin response at low
energies and long wavelengths (see, e.g.,
Refs.~\cite{Lykasov:2005,Lykasov:2008}), whereas in the case of
central interactions or at the single-nucleon level, bremsstrahlung is
forbidden by conservation laws.

The case of nucleon-nucleon (NN) bremsstrahlung is rather well
studied. The tensor part of the leading one-pion-exchange interaction
gives rise to NN bremsstrahlung, which was first calculated in the
pioneering work of Friman and Maxwell for degenerate conditions in
neutron star cooling~\cite{Friman:1979} and developed into a structure
factor for general conditions in supernova simulations by Hannestad
and Raffelt~\cite{Hannestad:1998}. For NN scattering, there are
important contributions beyond one-pion-exchange, which have been
calculated based on NN phase shifts~\cite{Hanhart:2001,Bacca:2012} and
in chiral effective field theory~\cite{Bacca:2009,Bacca:2012}.

The presence of nuclei can provide additional contributions to the
spin structure factor, thus increasing the spin relaxation rate. In
this paper, we consider the contributions from $\alpha$ particles to
nucleon bremsstrahlung processes. Motivated by Ref.~\cite{Bacca:2012},
we will focus on non-degenerate conditions. Nucleon-$\alpha$
scattering features a P-wave resonance near $1$~MeV, which can be seen
as a single-particle excitation on top of a $\alpha$ core where the
S-wave states are filled. The spin-orbit interaction splits the
$^{2}\pw_{{3}/{2}}$ and the $^{2}\pw_{{1}/{2}}$ waves, and hence this
channel contributes to the spin structure factor.

In this paper, we will focus on the comparison between bremsstrahlung
in neutron-neutron ($nn$) and $n\alpha$ scattering. We calculate the
$n\alpha$ contribution and point out regimes where this scattering
process is competitive with $nn$ scattering in the production of
neutrino pairs. Our main results are summarized in Figs.~\ref{fig:Xi}
and~\ref{fig:T5emissivity_rho_vs_T}, which show that for equal number
densities of $\alpha$ and $n$, and for temperatures $T \lesssim4$~MeV,
the $n\alpha$ contribution to the spin structure factor is
significantly larger than the $nn$ one.

\section{Neutron-$\alpha$ bremsstrahlung}

\begin{figure}[t]
\begin{center}
\includegraphics[width=0.95\columnwidth,clip=]{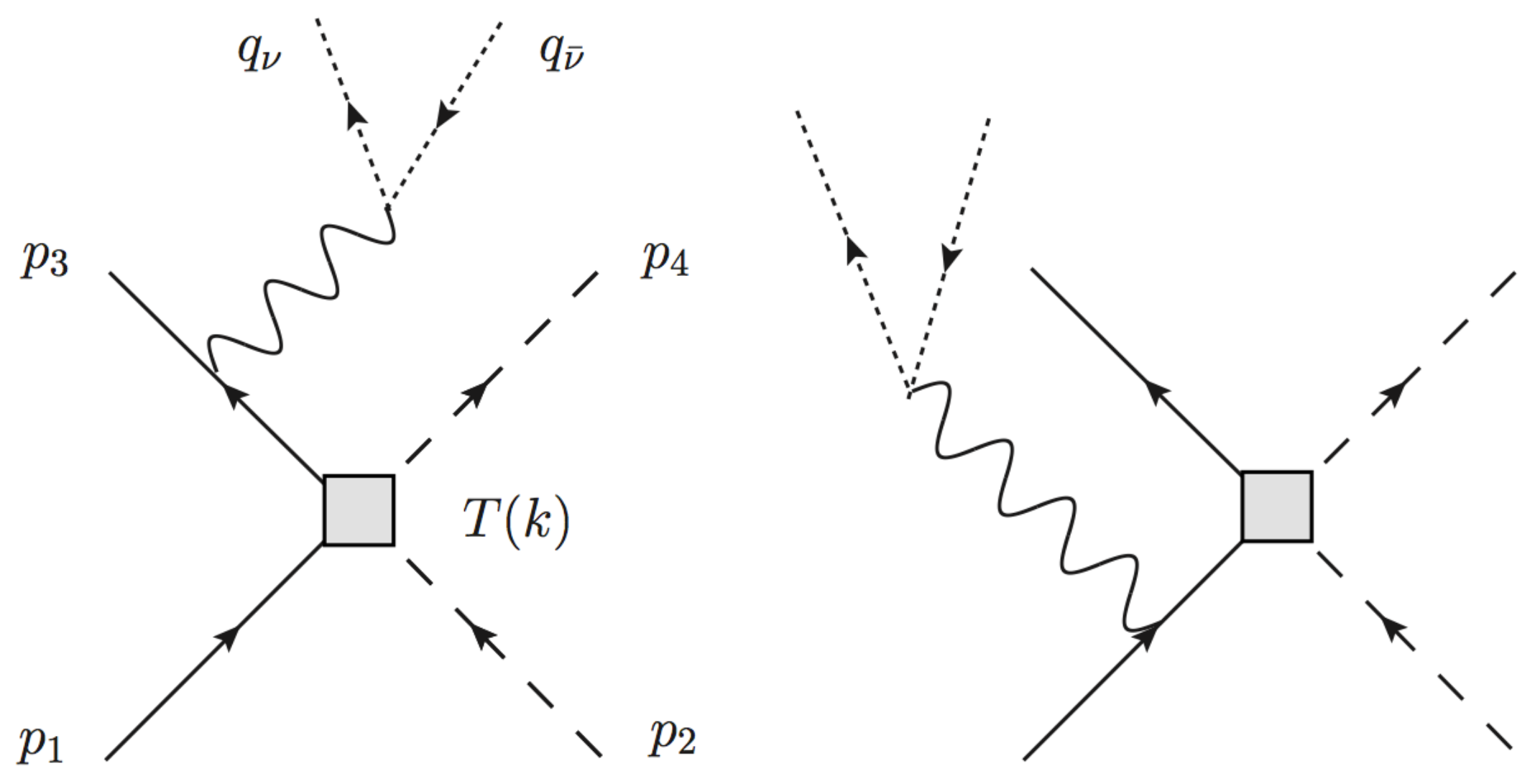}
\end{center}
\caption{Diagrams contributing to the $n\alpha \longleftrightarrow 
n\alpha \nu\bnu$ process. The solid line denotes neutrons $n$, the
dashed lines $\alpha$ particles, the dotted line the $\nu\bar{\nu}$
pair, and the wiggly line is a $Z^0$ boson exchange. For $nn$
bremsstrahlung, the $\nu\bar{\nu}$ pair can also be radiated from the
second neutron.\label{fig:BasicBremsstrahlung}}
\end{figure}

We consider $\nu\bnu$ bremsstrahlung from $n\alpha$ scattering shown
diagrammtically in Fig.~\ref{fig:BasicBremsstrahlung}. The incoming
four-momenta of the two hadronic particles are denoted by $p_1$ and
$p_2$, while their final momenta are $p_3$ and $p_4$. Because $\alpha$
particles are spin-less, only the neutron radiates a $\nu\bnu$ pair
(with four-momenta $q_\nu$ and $q_{\bnu}$) via the exchange of a
$Z^0$ boson. The scattering amplitude $\calM$ for this process can be
written as
\begin{equation}
i \calM = \frac{i \, G_F C_A}{\sqrt{2}} \frac{1}{-\omega} 
\sum_{j=1,2,3} l^j
 \chi_1^\dagger \, [{\boldsymbol \sigma}^j , \calT({\bf k}) ] \,
\chi_3 \,, \label{eq:generalM}
\end{equation}
where $G_F$ is the Fermi coupling constant and $C_A=-g_A/2=-1.26/2$ is
the axial-vector coupling for neutrons. Here, $\omega = - (q^0_\nu +
q^0_{\bnu})$ is the energy transferred from the neutrino pair to the
neutron, and ${\bf k} = {\bf p}_3-{\bf q}-{\bf p}_1$ is the momentum
transfer, with ${\bf q}=-({\bf q}_\nu+{\bf q}_{\bnu})$. Moreover,
$l^j$ is the leptonic current, ${\boldsymbol \sigma}^j$ are Pauli
matrices associated with the neutron spin, $\chi_{1,3}$ are neutron
spinors, and $\calT({\bf k})$ denotes the $n\alpha$ scattering
vertex. Note that for $nn$ bremsstrahlung there are two additional
diagrams associated with the $Z^0$ boson emitted from the $2$ and $4$
neutron, in addition to the exchange diagrams~\cite{Friman:1979}.

To simplify the calculation, we approximate $q = |{\bf q}| \approx 0$
so that ${\bf k} \approx {\bf p}_3 -{\bf p}_1$. This is justified
because $|{\bf q}_\nu+{\bf q}_{\bnu}|$ is of the order of the
temperature $T$, which is small compared to the neutron momenta. In
fact, the magnitude of ${\bf p}_1$ and ${\bf p}_3$ are of the order of
the Fermi momentum $p_{\rm F}$ in the degenerate limit, or $\sqrt{2
m_N T}$ in the Boltzmann limit (with nucleon mass $m_N$), both of
which are greater than the typical temperature $T$. Since $\calM$ is
related to the commutator of the spin matrices with the scattering
vertex $\calT({\bf k})$, $n\alpha$ scattering is significant if it
has a non-central spin structure, and if it is not negligible compared
to $nn$ scattering.  Therefore, we next consider the $n\alpha$
scattering vertex and compare it with the $nn$ case.

\subsection{Neutron-$\alpha$ scattering vertex}

For $n\alpha$ scattering, the non-central structure arises from the
spin-orbit interaction, which leads to a splitting of the $\pw$ (and
higher) partial waves. As shown in ab initio calculations, this
spin-orbit splitting results from the spin-orbit NN interaction and
from non-central NN and 3N forces~\cite{Nollett:2007,Hupin:2013}.  The
$\sw$-wave channel gives zero contribution to $\calM$, because it
commutes with the spin. In the $\pw$-wave channels, the scattering
between $n$ and $\alpha$ is enhanced for relative momenta $p \sim
35$~MeV corresponding to the $^2\pw_{3/2}$ resonance seen at
laboratory energies\footnote{The relation between $E$ and the relative
momentum $p$ is given by $E=p^2 \, m_N/(2 \mred^2)$, where $\mred$ is the
reduced mass.}  $E \sim 1$~MeV.

\begin{figure}
\begin{center}
\includegraphics[width=0.95\columnwidth,clip=]{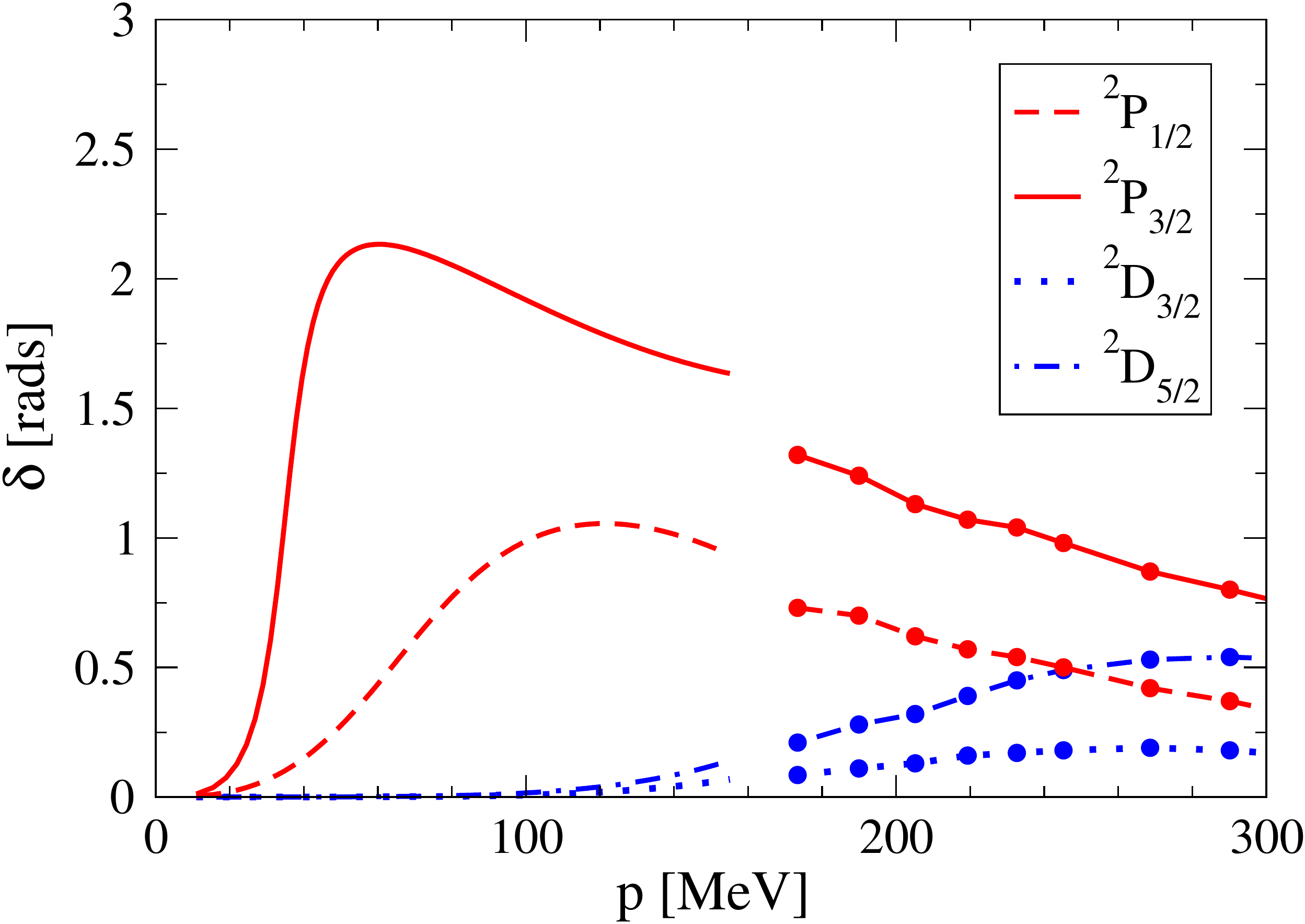}
\end{center}
\caption{(Color online) Phase shifts $\delta$ as a function of relative
momentum $p$ for $n\alpha$ scattering. The different partial waves are 
labelled by the standard notation $^{2S+1}\ell_j$. Phase
shifts are shown following Ref.~\cite{Horowitz:2006}: for low momenta
from the fits to data of Ref.~\cite{Arndt:1970} and for higher momenta (lines
with points) from optical model calculations~\cite{Amos:2000}. We do
not show the $^2\sw_{1/2}$ phase shift, because the corresponding $\calT$
matrix commutes with the spin operator in Eq.~(\ref{eq:generalM}) and
therefore, this channel does not contribute to $n\alpha$
bremsstrahlung.\label{fig:delta_nalpha}}
\end{figure}

This can be seen most clearly from the phase shifts shown in
Fig.~\ref{fig:delta_nalpha}. For $n\alpha$ scattering, where coupled
channels do not exist because the $\alpha$ particle has spin zero, the
commutator in Eq.~(\ref{eq:generalM}) requires the phase shifts for
the same $\ell$ and $j=\ell\pm1/2$ to be different. This effect is
largest in the $^2\pw_{3/2}$ and $^2\pw_{1/2}$ channels for $p \sim
50$~MeV due to the $\pw$-wave resonances. In contrast, the magnitudes
of the spin-1 (odd $\ell$) NN phase shifts that contribute to $nn$
bremsstrahlung are smaller than $0.2$~radians for $p \lesssim
200$~MeV~\cite{NNonline}. Therefore, we expect an enhancement for
$n\alpha$ bremsstrahlung in this momentum regime compared to the $nn$ case.

Because we consider the regime where matter is non-degenerate, the
$n\alpha$ scattering vertex relevant for the calculation is the
$\calT$ matrix~\cite{Bacca:2012}. We use the following conventions for
the definition of the $\calT$ matrix in terms of $n\alpha$ phase
shifts $\delta(p,\ell,S,j)$,
\begin{equation}
\langle p \ell S j| {\cal T} | p \ell S j \rangle =
\frac{1}{2\mred} \frac{1- e^{2i\delta(p,\ell,S,j)}}{2ip} \,,
\label{Eq:tmatrix}
\end{equation}
where $\ell, S$, and $j$ are the relative orbital angular momentum,
the neutron spin $S=1/2$, and the total angular momentum,
respectively. The initial and final relative momenta are given by
${\bf p}_i=\mred({\bf p}_1/m_N-{\bf p}_2/m_\alpha)$ and ${\bf p}_f
=\mred({\bf p}_3/m_N-{\bf p}_4/m_\alpha)$, where $\mred$ is the
reduced mass. Since the $\nu\bnu$ pair transfers the energy
$\omega$ to the neutron, 
we have $|{\bf p}_f|=\sqrt{{\bf p}_i^2+2\mred\omega}$.

For elastic scattering, the phase shifts are given in terms of the
on-shell momentum $p=|{\bf p}_i|=|{\bf p}_f|$. Following
Ref.~\cite{Bacca:2012}, we will approximate the scattering vertex for
bremsstrahlung by its on-shell value at the average of the initial and
final relative momenta, $p \approx (|{\bf p}_i|+|{\bf p}_f|)/2$. Note
that the momentum transfer ${\bf k} \approx {\bf p}_3-{\bf p}_1 = 
{\bf p}_f-{\bf p}_i$ is the same in the center-of-mass or the rest frame.

\subsection{Spin-summed matrix element}

For low neutrino momenta $q \ll p_{i,f}$, the spin-summed square of
the scattering amplitude can be expressed in a simple
form~\cite{Friman:1979}
\begin{equation}
\sum\limits_{\rm spins} |\calM|^2 
= 96 \, \frac{G_F^2 C_A^2}{2} \frac{1}{\omega^2} \, q_\nu^0 q_{\bnu}^0
\, W(p_i,p_f,\Theta) \,,
\end{equation} 
where $W(p_i,p_f,\Theta)$ is related to the hadronic part of the
scattering diagram~\cite{Bacca:2012}. For unpolarized initial states
and at the $\calT$ matrix level, the hadronic part depends only on the
magnitude of the initial and final relative momenta and on the
scattering angle $\cos \Theta = \widehat{\bf p}_i \cdot \widehat{\bf p}_f$.

For the $nn$ case, $W_{nn}$ is given by~\cite{Bacca:2012}
\begin{align}
& W_{nn}(p_i,p_f,\Theta) = \nonumber \\[1mm]
& \frac{1}{12} \sum_{j=1,2,3} {\rm Tr}
\Bigl\{ \left( \langle 34 | \calT({\bf k})| 12\rangle^* 
-\langle 43 |\calT'({\bf k}') |12\rangle^* \right) 
{\boldsymbol \sigma}_1^j \nonumber \\
&\quad \Bigl[ 
({\boldsymbol \sigma}_1+{\boldsymbol \sigma}_2)^j,
\left( \langle 34 | \calT({\bf k})|12 \rangle 
- \langle 43 |\calT'({\bf k}') | 12\rangle \right) \Bigr] \Bigr\} \,,
\end{align}
where ${\bf k}' \approx {\bf p}_4-{\bf p}_1$, and the terms
$\calT'({\bf k}')$ arise from the exchange diagrams. The initial state
is written as $|12\rangle$, which is shorthand for the initial
momentum and spins $|12\rangle=|{\bf p}_1 s_1 {\bf p}_2 s_2
\rangle$. Similarly, $\langle34|$ refers to the final state.

For the $n\alpha$ case, where no exchange diagrams are present,
$W_{n\alpha}$ becomes
\begin{align}
&W_{n\alpha}(p_i,p_f,\Theta) = \nonumber \\[1mm]
&\frac{1}{12} \sum_{j=1,2,3} {\rm Tr} \Bigl\{ 
\langle 34|(\calT({\bf k}))| 12\rangle^* {\boldsymbol \sigma}_1^j
\, \Bigl[ {\boldsymbol \sigma}_1^j, \langle 34| \calT({\bf k})|12 \rangle
\Bigr] \Bigr\} \,.
\end{align}

\subsection{Phase-space integral,\\ 
emissivity and spin structure factor}

The rate of production of $\nu\bnu$ pairs with energy $-\omega$ is
obtained by integrating $\sum |\calM|^2$ over the phase space. The
emissivity $\varepsilon_{\nu\bnu}(-\omega)$, i.e., the number of
$\nu\bnu$ pairs emitted per unit time and unit mass of matter, is
given by
\begin{align}
&\varepsilon_{\nu\bnu}(-\omega) =
\frac{\zeta_n\zeta_X}{\rho} \int \frac{d^3 q_\nu}{2q^0_\nu \, (2\pi)^3}
\frac{d^3 q_{\bnu}}{2q^0_{\bnu} \, (2\pi)^3} \,
\delta(q_\nu^0+q_{\bnu}^0+\omega) \nonumber \\[1mm]
& \int \frac{d^3 p_1}{(2\pi)^3} \frac{d^3 p_2}{(2\pi)^3}
\frac{d^3 p_3}{(2\pi)^3} \frac{d^3 p_4}{(2\pi)^3} \,
\bigl[n_1 n_2 (1-n_3) (1 \pm n_4)\bigr] \nonumber \\[1mm]
& \frac{1}{f_s} \sum\limits_{\rm spins} |\calM|^2 \,
(2\pi)^4 \delta(q_\nu^\mu+q_{\bnu}^\mu+p_3^\mu+p_4^\mu-p_1^\mu-p_2^\mu) \,,
\label{eq:rate}
\end{align}
where $\rho$ is the mass density, $\zeta_X$ are the spin degeneracies
with $X=n$ or $\alpha$ ($\zeta_n=2$, $\zeta_\alpha=1$), $n_{1,3} =
1/[\exp((\epsilon_{1,3}-\mu)/T)+1]$ are Fermi distribution
functions for the neutrons and $n_{2,4} =
1/[\exp((\epsilon_{1,3}-\mu)/T)\pm 1]$ are Fermi or Bose
distribution functions for the species $X$ corresponding to its
statistics. The symmetry factor $f_s$ is $4$ when the scattering
particles are identical (a factor of $2$ for the initial and final
states each), and $1$ otherwise. The sign $\pm$ in Eq.~(\ref{eq:rate})
is negative for fermions ($n$) and positive for bosons ($\alpha$).

It is useful to write Eq.~(\ref{eq:rate}) in terms of the spin
structure factor $S_\sigma(\omega,{\bf q})$~\cite{RaffeltBook}
(where we use the same convention as Ref.~\cite{Bacca:2012})
\begin{equation}
S_\sigma(\omega,{\bf q}) 
= \frac{1}{\pi n_n} \frac{1}{1-e^{-\omega/T}} \, {\rm Im} 
\chi_{\sigma}(\omega,{\bf q}) \,, \label{eq:S_sigma}
\end{equation}
where $\chi_{\sigma}(\omega,{\bf q})$ is the spin response
function. All neutrino processes (scattering, emission, and
absorption) are determined by the spin structure factor. For example,
the emissivity is given by~\cite{RaffeltBook}
\begin{equation}
\varepsilon_{\nu\bnu}(-\omega) = \frac{n_n}{\rho} \, G_F^2 C_A^2 \,
\frac{\omega^5}{20\pi^3} \, S_\sigma(-\omega,q=0) \,,
\end{equation}
where we have taken the long-wavelength limit for the neutrinos.
Detailed balance implies that $S_\sigma(-\omega,q=0) = e^{-\omega/T}
S_{\sigma}(\omega,q=0)$. The total energy-loss rate $Q_{\nu\bnu}$ per
unit time and unit mass of matter is then given by
\begin{equation}
Q_{\nu\bnu} = \int_0^\infty d\omega \, \omega \, \varepsilon_{\nu\bnu}(\omega) \,.
\end{equation}

For non-degenerate conditions, we can write the spin structure factor
in the long-wavelength limit as~\cite{Bacca:2012}
\begin{equation}
S_{\sigma}(\omega,q=0) = \frac{2}{\pi^2} \frac{n_X \, \mred}{f_s \, \omega T}
\frac{1}{1-e^{-\omega/T}} \, \Xi(\omega) \,, 
\label{eq:S_simplified}
\end{equation}
with the function
\begin{equation}
\Xi(\omega) = \frac{2 \sinh[\omega/(2T)]}{\omega/T} \,
\bigl\langle (p_i^2 + 2\mred\omega)^{1/2} \, W \, e^{-\omega/(2T)} 
\bigr\rangle \,. \label{eq:Xi}
\end{equation}
The advantage of using $\Xi(\omega)$ is that it is independent of the
density of neutrons and $\alpha$ particles and depends only on the
energy transfer and the temperature. Therefore, it allows us to
compare the contributions from $n\alpha$ and $nn$ scattering,
considering only the strength of the interactions, without
needing to specify their densities. In Eq.~(\ref{eq:Xi}),
$\langle \, \cdots \rangle$ stands for the average with a
Boltzmann weight~\cite{Bacca:2012},
\begin{align}
\langle \, \dots \rangle &= \frac{\int_0^{\infty} dp_i \, p_i^2 \, 
e^{-p_i^2/(2\mred T)} \int d\cos\Theta \, \dots \,}{\int_0^{\infty} 
dp_i \, p_i^2 \, e^{-p_i^2/(2\mred T)} \int d\cos\Theta} \,, \nonumber \\[1mm]
&= \frac{\int_0^{\infty} dp_i \, p_i^2 \, e^{-p_i^2/(2\mred T)} 
\int d\cos\Theta \, \dots \,}{(2\mred T)^{3/2} \, \Gamma(3/2)} \,.
\end{align}

Using angular momentum algebra, one can express $W$ as a sum over
partial-wave contributions, analytically integrate over $\Theta$, and
hence obtain an expression $\Xi(\omega)$ in terms of the matrix
elements $\langle p \ell S j | \calT | p \ell S j \rangle$ of
Eq.~(\ref{Eq:tmatrix}). For the $nn$ case, the expression for
$\Xi(\omega)$ is given by Eq.~(43) in Ref.~\cite{Bacca:2012}. The
$n\alpha$ case results in a similar expression:
\begin{align}
&\Xi(\omega) = \frac{2}{\sqrt{\pi} (2m_{\rm{red}} T)^{3/2}} \,
\frac{2\sinh[\omega/(2T)]}{\omega/T} 
\, \frac{(4 \pi)^2}{2} \nonumber\\
&\times \sum_{\ell>0 \, L} \, \sum_{j \tilde{j}} \, \sum_{m_S m'_S}
(-1)^{j+\tilde{j}+L} \, \bigl( \, \widehat{j} \, \widehat{\tilde{j}} \,
\widehat{L} \, \bigr)^2 \, \widehat{\ell}^2 \, \nonumber \\
&\times \left\{
\begin{array}{c c c}
\ell & \ell & L \\
1/2 & 1/2 & j \\
\end{array}
\right\}
\left\{
\begin{array}{c c c}
\ell & \ell & L \\
1/2 & 1/2 & \tilde{j} \\
\end{array}
\right\}
\left\{
\begin{array}{c c c}
\ell & \ell & L \\
\ell & \ell & 0 \\
\end{array}
\right\} \nonumber \\
&\times \biggl[ {\cal C}_{L (m_S-m'_S)1/2\,m'_S}^{1/2\,m_S} \biggr]^{2}
(m_S^{2}-m_S m'_S) \nonumber \\
&\times \int_0^\infty dp_i \, p_i^2 \, \sqrt{p_i^2+2m_{\rm{red}}\omega} 
\, e^{-p_i^2/(2m_{\rm{red}}T)-\omega/(2T)} \nonumber \\
&\times \bigl\langle \sqrt{p_i^2+2m_{\rm{red}}\omega} \, | \calT_{\ell S j}
| p_i \bigr\rangle
\bigl\langle \sqrt{p_i^2+2m_{\rm{red}}\omega} \, | \calT_{\ell S \tilde{j}}
|p_i \bigr\rangle \,,
\label{pwformula}
\end{align}
with $m_S, m_S' = \pm 1/2$, $\widehat{a}=\sqrt{2a+1}$ and standard
notation for the Clebsch-Gordan, 3j, and 6j symbols (see
Ref.~\cite{Bacca:2012}).

\section{Results}

\begin{figure}[t]
\begin{center}
\includegraphics[width=0.95\columnwidth,clip=]{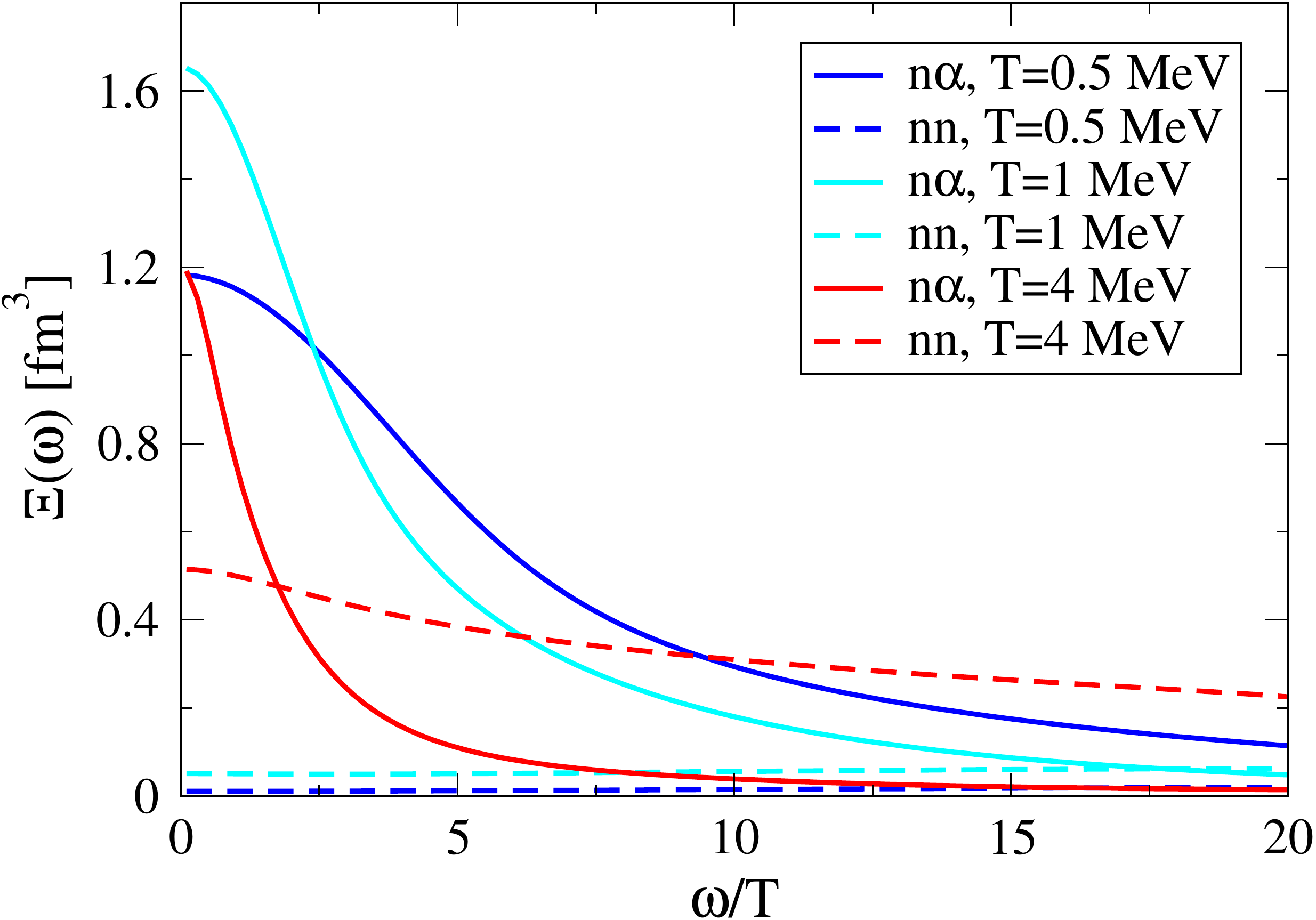}
\end{center}
\caption{(Color online) $\Xi(\omega)$ as a function of $\omega/T$ for 
various temperatures. The solid lines are for $n\alpha$, while the 
dashed lines are for $nn$. For fixed $\omega/T$, $\Xi(\omega)$ for
$n\alpha$ bremsstrahlung decreases with increasing $T$ (except for 
low $\omega/T$) and increases for the $nn$ case.\label{fig:Xi}}
\end{figure}

For $n\alpha$ scattering, the $\pw$-wave resonances lead to a peak in
the hadronic trace $W_{n\alpha}(p_i,p_f,\Theta)$ at $p \sim 40$~MeV
which drops off for $p \gtrsim100$~MeV.  This is to be contrasted with
the $nn$ contribution. Because $W_{nn}$ increases monotonically with
relative momentum up to $p \sim 150$~MeV, increasing $T$ (or $\omega$)
leads to an increased response.  Therefore, we expect that the spin
response for $n\alpha$ scattering should be large for $T$ less than a
few MeV, while $nn$ should dominate at higher $T$. The detailed
calculation for $\Xi(\omega)$ shows that this is indeed the case. In
Fig.~\ref{fig:Xi}, we observe that for $T=0.5$~MeV and $T=1$~MeV,
$\Xi(\omega)$ for $n\alpha$ dominates over $nn$ and it is larger by
several orders of magnitude. As we increase $T$, the $n\alpha$
response decreases and the $nn$ response increases. But even for
$T=4$~MeV, the $n\alpha$ response is larger for $\omega/T < 2$.
Consequently, we expect that the neutrino-pair production and
absorption could be affected by $n\alpha$ processes if
$T\lesssim4$~MeV and the density of $\alpha$ particles is not orders
of magnitude smaller than the neutron density.

\begin{figure}[t]
\begin{center}
\includegraphics[width=0.95\columnwidth,clip=]{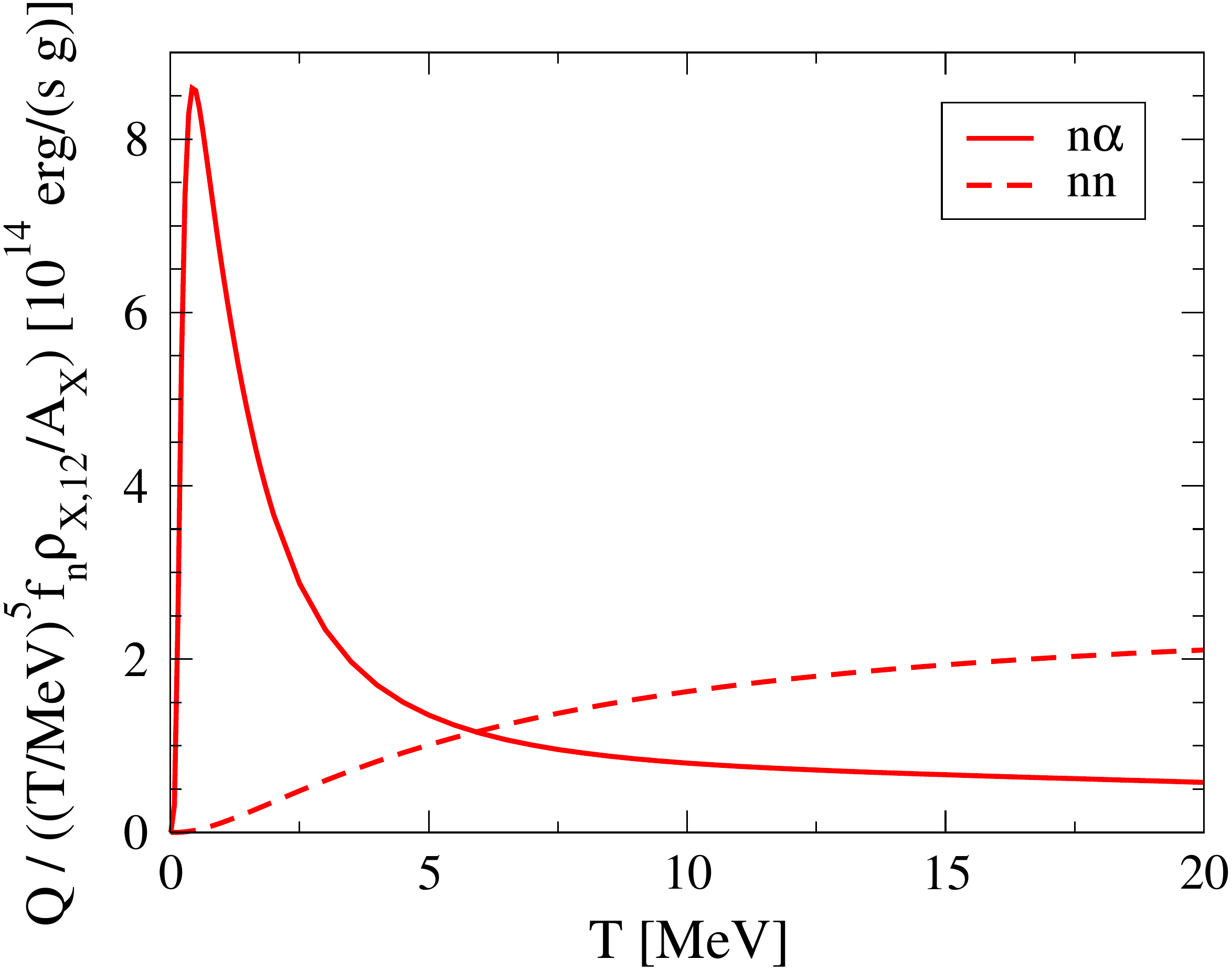}
\end{center}
\caption{(Color online) Energy-loss rates $Q_{\nu\bnu}$ as a function
of temperature $T$. The solid (dashed) line is for $n\alpha$ ($nn$)
bremsstrahlung. The $Q_{\nu\bnu}$ shown is divided by $T^5$ (in MeV),
by the neutron mass fraction $f_n$, and by the mass density over the mass
number $\rho_{X,12}/A_X$ of $X=n$ or $\alpha$ particles (in
$10^{12}$~g\,cm$^{-3}$), with $A_X=1$ or $4$,
respectively.\label{fig:T5emissivity_rho_vs_T}}
\end{figure}

From the resulting $\Xi(\omega)$ and $S_\sigma(\omega)$, one can
readily calculate neutrino rates. In
Fig.~\ref{fig:T5emissivity_rho_vs_T}, we show the behavior of the
energy-loss rates $Q_{\nu\bnu}$ as a function of temperature. The rate
for $nn$ bremsstrahlung is based on NN phase shifts (given by the
${\cal{T}}$~matrix in Ref.~\cite{Bacca:2012}). We observe that, for the same
neutron mass fraction $f_n$ and if the $\alpha$-particle density
$\rho_\alpha/4 = m_N \, n_\alpha$ is comparable to $\rho_n = m_N \,
n_n$, $n\alpha$ bremsstrahlung dominates over $nn$ for $T <
6$~MeV. Even if $\rho_\alpha/4$ is much smaller than $\rho_{12}^n$,
$n\alpha$ scattering could be the dominant process for smaller
$T$. For example, for $T=2$~MeV, the ratio of the energy-loss rates
(with the densities scaled out) is $\sim 10$ and increases as
we further decrease the temperature.

Next, we provide simple expressions that characterize the energy-loss
rates $Q_{\nu\bnu}$ shown in Fig.~\ref{fig:T5emissivity_rho_vs_T}. For
$n\alpha$ bremsstrahlung, the energy-loss rate [in erg/(s\,g)] is
given by the fit function
\begin{equation}
Q_{\nu\bnu}^{n\alpha} = \frac{8.2 \cdot 10^{14} \, T^2}{(T+0.07)^3 + 0.4^3}
\, T^5 \, f_n \, \rho_{\alpha,12}/4 \,,
\end{equation}
and for $nn$ bremsstrahlung, we have
\begin{equation}
Q_{\nu\bnu}^{nn} = \bigl(7.4 \cdot 10^{12} \, T + 8.4 \cdot 10^{13}\bigr) \,
\frac{T^3}{T^3+2.4^3} \, T^5 \, f_n \, \rho_{n,12} \,,
\end{equation}
where temperatures $T$ are expressed in MeV. These parameterizations
hold for non-degenerate conditions and can be used not only to
compare the $nn$ and $n\alpha$ rates to each other, but also to other
competing processes.

So far, we have emphasized the role of the $\pw$-wave resonances for
the $n\alpha$ spin response. The $\dw$-wave channel also has a
spin-orbit splitting and contributes to $W_{n\alpha}$.  From
Fig.~\ref{fig:delta_nalpha}, we expect the $\dw$-wave contribution to
be smaller than the $\pw$-wave contribution for relative momenta $p <
250$~MeV. We find that this is indeed the case. For the temperatures
of interest the $\dw$-wave contribution can be neglected, although we
have included it in the present calculation. It only shows up at
high momenta where the Boltzmann factors are small.

We also comment on the role of inelastic channels in $n\alpha$
scattering. For energies greater than $20$~MeV (in the
$\alpha$-particle rest frame), corresponding to $p \gtrsim155$~MeV,
$n\alpha$ scattering has inelastic channels. These can be
parameterized by an imaginary part in the phase shifts, as was done in
the optical model calculations of Ref.~\cite{Amos:2000}. However, the
Boltzmann factors for such large energy transfers are small, making
the effects of inelastic channels negligible for $T \lesssim 4$~MeV.

\section{Conclusions}

Our main conclusion is that the $\pw$-wave resonances in $n\alpha$
scattering lead to an enhanced contribution to the nucleon spin
structure factor for temperatures $T \lesssim 4$~MeV. The contribution
from $n\alpha$ scattering per $\alpha$ particle is orders of magnitude
larger than the $nn$ contribution per neutron for $T<1$~MeV, and for
these temperatures, may be relevant even if the number density of
$\alpha$ particles is much smaller than of nucleons. As we increase
$T$, the $nn$ contribution increases and the $n\alpha$ contribution
decreases but even up to $T=4$~MeV, the $n\alpha$ contribution is
larger for $\omega/T < 2$. Since the spin structure factor is directly
related to neutrino processes in nuclear matter, this may impact
neutrino production and propagation in environments with substantial
$\alpha$ particles and nucleons. As an example we considered neutrino
emissivities, and found that the $n\alpha$ contribution to the
energy-loss rate per $\alpha$ particle is larger than the $nn$
contribution per neutron for $T<6$~MeV.

Recently, a resonant enhancement due to large $\sw$-wave scattering
lengths was found for NN bremsstrahlung at densities $\rho \lesssim
10^{13}$~g\,cm$^{-3}$ when protons are present~\cite{Bartl:2014}.
Note that we have compared to $nn$ bremsstrahlung, where there is no
low-density enhancement.

In conclusion, $n\alpha$ bremsstrahlung processes can be competitive
with other neutrino emission processes in environments where both
$\alpha$ particles and nucleons are abundant, and temperatures are in
the range $0.1-4$~MeV. For example, the outer layers of proto-neutron
stars feature neutrons, protons, and $\alpha$ particles (e.g., see the
equations of states used in Ref.~\cite{Arcones:2008}).  These can
have observable implications on the spectra and the flux of
neutrinos~\cite{Arcones:2008} coming from core-collapse supernovae
both from a modification in opacity and production.

These processes may also play a role in cores of giant stars when they
are rich in $\alpha$ particles. Our calculations of the energy-loss rate and the spin
structue factor, Eq.~(\ref{eq:S_simplified}) combined with
Fig.~\ref{fig:Xi}, can be used to investigate the contribution of
$n\alpha$ bremsstrahlung processes as a function of $T$ and to compare
with electronic processes that dominate neutrino production in these stars.

Finally, we comment about the broader implications of nucleon-nucleus
processes, where similar enhancements of neutrino-pair bremsstrahlung
are expected for resonant non-central interactions. Note that the same
non-central physics is at play in the recent finding~\cite{Misch:2013}
of an increased $\nu\bnu$ emission rate from thermally excited nuclei
due to spin-orbit splittings. Therefore, even if $\alpha$ particles
are not abundant, resonant non-central nucleon-nucleus scattering can
lead to enhanced neutrino emission.

\begin{acknowledgements}

We thank Sanjay Reddy for discussions. This work was supported in part
by the Natural Sciences and Engineering Research Council (NSERC), the
National Research Council of Canada, the ERC Grant No.~307986
STRONGINT, and the Helmholtz Alliance Program of the Helmholtz
Association contract HA216/EMMI ``Extremes of Density and Temperature:
Cosmic Matter in the Laboratory''.

\end{acknowledgements}

\bibliography{local}

\end{document}